\documentstyle	[12pt]{article}

\let\a=\alpha	\let\b=\beta	\let\g=\gamma
\let\d=\delta	\let\e=\epsilon 
	\let\q=\theta	
\let\k=\kappa 	\let\m=\mu
\let\n=\nu			
\let\s=\sigma		
\let\f=\phi	\let\c=\chi	\let\y=\psi
	\let\G=\Gamma

\let\F=\Phi	\let\Y=\Psi 
 \let\la=\label	\let\ci=\cite

\def\bd{\begin{document}}
\def\ed{\end{document}}
\def\ds{\documentstyle}	\let\fr=\frac
\let\bl=\bigl	\let\br=\bigr \let\Br=\Bigr
\let\Bl=\Bigl \let\bm=\bibitem
\let\na=\nabla \let\pa=\partial
\let\ov=\overline
\newcommand{\be}{\begin{equation}}
\newcommand{\ee}{\end{equation}}
\def\ba{\begin{array}} \def\ea{\end{array}}
\newcommand{\ho}[1]{$\,	^{#1}$}
\newcommand{\hoch}[1]{$\,	^{#1}$}
\newcommand{\bea}{\begin{eqnarray}}
\newcommand{\eea}{\end{eqnarray}}
\newcommand{\ra}{\rightarrow}
\newcommand{\lra}{\longrightarrow}
\newcommand{\Lra}{\Leftrightarrow}
\newcommand{\ap}{\alpha^\prime}
\newcommand{\bp}{\beta^\prime}
\newcommand{\tr}{{\rm	tr}	}
\newcommand{\Tr}{{\rm	Tr}	}
\newcommand{\NP}{Nucl.	Phys.	}
\newcommand{\tamphys}{\it	Center	for
Theoretical	Physics\\ Physics	Department	\\
Texas	A	\&	M	University \\	College	Station,
Texas	77843}

\begin{document}

\hfill{CTP-TAMU-45/93}

\vspace{12pt}

\hfill{hep-th/9308088}

\vspace{24pt}

\begin{center} { \large {\bf The Search for Supersymmetry
Anomalies--Does Supersymmetry Break Itself?}}

\vspace{12pt}

{\bf Talk given at the HARC conference \\
on \\
`Recent Advances in the Superworld'}

\vspace{36pt}

J. A. Dixon\footnote{Supported in part by the U.S. Dept of
Energy, under grant DE-FG05-91ER40633, Email: dixon@phys.tamu.edu}

\vspace{6pt}

{\tamphys}

\vspace{6pt}

August, 1993

\vspace{6pt}

\underline{ABSTRACT}

\end{center}

The established results concerning the BRS cohomology
of supersymmetric theories in four space-time dimensions are
briefly reviewed.  The current status of knowledge concerning
supersymmetry anomalies and the possibility that supersymmetry
breaks itself through anomalies in local composite operators
is then discussed.

It turns out that the simplest allowable
supersymmetry anomalies occur only in conjunction with
the spontaneous breaking of gauge symmetry.  A simple example
of such a possible supersymmetry anomaly is presented.

\vfill

\baselineskip=24pt

\pagebreak

\setcounter{page}{1}

\section{Review and Discussion}

Supersymmetry is currently one of the most
popular ways to extend the standard model of
strong, weak and electromagnetic
interactions.   It is also an essential
ingredient of the superstring which is the
only known candidate for a theory that might
include gravity in a consistent way.
However a major difficulty of supersymmetry
is to explain  why  we do not observe it,
assuming that it is really present. How and
why is supersymmetry broken? It is possible
that the considerations of the present paper
may be relevant to that question.

It has been known for some time now that the
BRS cohomology of  $N=1,D=4$ rigid
supersymmetry is very complicated \ci{cmp2}\ci{cqg}\ci{prl}.
All the non-trivial structure occurs in Lorentz non-trivial
sectors, and was consequently not found in some
investigations, for example \ci{bdk}, but see also
\ci{b1}\ci{b2}.

This means that in principle there may be
anomalies in these theories that violate
supersymmetry.  We know that anomalies
usually play an important role in theories,
so any examples of such anomalies would be
interesting.

    In this paper, I briefly touch on some recent \ci{dm}\ci{dmr}
and also the older
results concerning
the BRS cohomology and then go on to some
remarks concerning the current state of research on these
questions \ci{pd1}\ci{pd2}.

The main problem now is that no supersymmetry anomalies have yet
been found that correspond to this
cohomology.  I shall try to explain some
of the fairly obvious reasons that such
anomalies are not present in some
simple cases, and then go on to exhibit a
case which appears to require a detailed
calculation to determine whether a
supersymmetry anomaly is or is not present.
Some of this material was not known at the time of the
talk given at the conference, but it seems appropriate
to include it here.

It is easy to write down the simplest
examples where a supersymmetry anomaly could
conceivably arise. The BRS cohomology of
these theories indicates that there could be
anomalies in the renormalization of
composite operators (made from the
elementary chiral superfields $S$ of the
theory) which are antichiral spinor
superfields.  These composite operators take forms such as: \be
\Y_{1 \a} = D^2 [ S_1 D_{\a} S_2 ]
\la{e1}
\ee \be
\Y_{2 \a} = {\ov S}_1 D^2 [ S_2  D_{\a} S_3 ]
\ee \be  \Y_{3 \a} = {\ov S}_1 D^2 [ {\ov
D}^2 {\ov S}_1   D_{\a} S_3 ]
\la{e3}
\ee
One could
add more chiral superfields $S$ or more
supercovariant derivatives of course.  The
main things to  keep in mind are:
\begin{enumerate} \item The expression for
$\Y_{1 \a}$ should not vanish
\item It is frequently necessary to use more
than one flavour of superfield $S$ to
prevent the expression from vanishing,
because such expressions may be
antisymmetric under interchange of flavour
indices \end{enumerate}

To calculate the anomaly, one would couple
such composite operators to the action with
an elementary (i.e. not composite)
antichiral spinor source superfield  $\F^{\a}$.
This means that one simply adds the following
term to the usual action of the theory:
\be
S_{\rm \F}  = \int d^6 {\ov z} \; \F^{\a}
\Y_{\a} \la{action}
\ee
where $\Y_{\a}$ is some composite antichiral
spinor superfield, examples of which are given above
in (\ref{e1}-\ref{e3}).

Then the anomaly would appear in the form:
\be  \d \G = m^k \int d^6 {\ov z} \; \F^{\a}
c_{\a} {\ov S}^n   \la{anom} \ee  where $\G$
is the one-particle irreducible generating
functional, $\d$ is the nilpotent BRS
operator, $\int d^6 {\ov z}$ is an integral
over antichiral superspace, $c_{\a}$ is the
constant ghost parameter of rigid
supersymmetry, $m^k$ is the  mass parameter
$m$ to some power $k$ required by simple
dimensional analysis, and  ${\ov S}^n $ is
the  $n^{th}$ power of the antichiral
superfield ( this might include  a sum over
indices which distinguish different $\ov S$
superfields from each other).

To count masses we use the following
assignments for the variables and the
superfields (Notation defined below):
\be
  m=1;\pa_{\m} =1; \q_{\a}=\fr{-1}{2}; S=1;
\F_{\a}= \fr{1}{2}; \ee
Now we define the component fields:
\be
S = A + \q^{\a} \y_{\a} + \fr{1}{2} \q^2 F
\ee
\be
\F_{\a} = \f_{\a} + W_{\a \dot \b} {\ov \q}^{\dot \b}
+ \fr{1}{2} {\ov \q}^2 \c_{\a}
\ee
The dimensions of these coefficient fields are then:
\[
   D_{\a}=\fr{-1}{2}; A=1; \y_{\a} =
\fr{3}{2}; F=2; \]
\be
  \c_{\a}=
\fr{3}{2};\f_{\a}= \fr{1}{2};W_{\a \dot \b}=
1;
\ee

An examination of  examples shows that
elementary dimensional  counting prevents
the powers of $m$ from working correctly to
yield (\ref{anom})
whenever the only vertices
of the diagram are chiral vertices involving
only chiral fields.  It should be possible to show
this by a dimensional argument, but I have not
taken the time to try to do that yet--at any rate it certainly
seems to hold for a wealth of examples, one of which can
be found in \ci{erice}.

  However when there is at
least one gauge propagator in the diagram,
the powers of $m$ easily work out correctly to
yield (\ref{anom}).   But then one has to
confront another problem, which is that one
has to analyze the cohomology again in the
presence of the gauge fields.
This unsolved problem has been partially finessed in
\ci{pd1}.

Another problem that is also unsolved is the
problem of solving the full BRS cohomology of
any supersymmetric theory including the
sources that are necessary to formulate the
full BRS identity.  Essentially, this brings
in the complication of ensuring that the BRS
cohomology space is orthogonal to the equations
of motion of the fields.  I will also make here some
new comments on this question, which has been the
subject of my work over the summer \ci{pd1}.

 When one tries to compute the full BRS cohomology of chiral
matter coupled to gauge fields, the  BRS
operator becomes quite formidable, and I will
not try to give all the details here.
However we can discuss a sub-problem of interest
without solving
the entire problem, and that is what we do
below.

Whenever one formulates a BRS identity in
the manner pioneered by Zinn-Justin, it is necessary to
also include sources ${\tilde f}_i$ for the
variation of the fields $f_i$, and in the
resulting `full' BRS operator, these give
rise to terms that involve the equations of
motion of the corresponding fields.    This
turns out to be  more or less equivalent to
the Batalin-Vilkovisky quantization method.
The essential point is that this eliminates
from the cohomology space anything which
vanishes by the equation of motion, i.e.
anything which vanishes `on-shell'.
In our case this will eliminate
all those objects in the cohomology space
which involve superfields $\ov S$ which have
mass terms in the action, as well as a
number of higher order terms that are of no
concern at present.

So we are now interested only in computing
diagrams where the possible supersymmetry
anomaly involves massless antichiral fields $\ov
S$ in (\ref{anom}).  But this raises another
problem.  The  most promising simple  case
(see below) seems to involve   a triangle diagram with
the $\F^{\a}$ superfield at one vertex, two
chiral (or antichiral) superfields emerging
from that vertex and the exchange of a
vector superfield between these two lines.
Now the mass counting implies that the
anomaly has a higher power of mass than the
composite operator from which it arises.
The only way this can happen is if some of
the interior lines are massive.  Is there
any  way that interior massive lines can
give rise to exterior massless lines while
exchanging a vector superfield? The answer to
this question is of course well known--this
will happen if and only if  the gauge
symmetry is spontaneously broken.  We will therefore
assume that gauge symmetry {\em is} spontaneously broken and that
supersymmetry {\em is not}
spontaneously broken. Since we are looking for supersymmetry
breaking through anomalies, it is reasonable to assume that
it is not otherwise broken.

 This combination is
in fact very easy to achieve--as is well
known, gauge symmetry breaking is natural
and very easy to achieve  in rigid
supersymmetry, but spontaneous supersymmetry
breaking can only be achieved with very
contrived models, particularly if the gauge
group is semisimple.

So now, if we want to examine the question
of supersymmetry anomalies, we are forced to
consider a supersymmetric gauge theory with
spontaneous breaking of the gauge symmetry.
But there are more conditions, at least for
the supersymmetry anomalies that involve
matter superfields.  In order for the
relevant diagrams to exist, we must have
matter multiplets which break under the gauge
breaking into a combination of massive and
massless fields, so that a massive vector superfield
can have a vertex with a massless and a massive chiral
superfield.

This happens of course for
the Higgs multiplet itself, but then the
massless Goldstone supermultiplets do not
contribute to the  BRS cohomology space, as
will be shown in the forthcoming paper \ci{pd1}.
We must have additional
(non-Higgs) matter multiplets which break
under the gauge breaking into a combination
of massive and massless fields.  There are
many ways to do this, and an example is
given below.  Note that this happens also in
the standard model, where  the neutrino
remains massless after spontaneous breaking
of $SU(2) \times U(1)$ to $  U(1)_{EM}$
simply  because there is no right handed
neutrino for it to  form a mass with, (and
because lepton conservation prevents the
formation of a Majorana neutrino mass, in
the minimal standard model at least). The
relevant discussion of the
standard model will also be the subject of a forthcoming
paper \ci{pd2}.

So if we want to find a supersymmetry anomaly,
we are driven to models with gauged
supersymmetry and spontaneous breaking of
the gauge symmetry through Higgs multiplets
which develop a VEV in their `A' components
(but not their `F' components--that would
break supersymmetry).  In addition these
models must have matter which is massless at
tree level, but which gets split into
massive and massless components as a result
of gauge breaking.    These are the only
models that have a chance of developing
supersymmetry anomalies in some of their
composite operators at the one loop level.  Such models are of
course  very  reminiscent of a
supersymmetric version of the standard model
of strong weak and electromagnetic
interactions.  It is just within the realm
of possibility that these anomalies could
account for the experimentally observed lack
of  supersymmetry  in the world with no
additional  assumptions in the model at
all--in which case we could say that
supersymmetry breaks itself.  But there is
plenty of work to do before we can determine
whether this notion is right.  Even if the supersymmetry anomalies
exist, considerable work will be necessary to deduce the
form of the
supersymmetry breaking they give rise to.

A rather interesting and new feature is that
we can see that the particular `soft'
mass-dependent supersymmetry anomalies we
are examining here, if there coefficients are non-zero, would
give rise to a kind of supersymmetry
breaking that is a function of the VEV
that    breaks the  gauge symmetries, and
which vanishes in the gauge symmetric
limit.     In addition,
it has been conjectured \ci{erice} that such anomalies might also provide a
natural mechanism whereby `supersymmetry breaks itself', while at
the same time retaining the cosmological constant at the zero value
it naturally has in many unbroken supersymmetric theories.
Clearly the spontaneous breaking of the gauge symmetry would
not interfere with this feature.

\section{A Simple Example}
\la{secexample1}
We consider a supersymmetric gauge theory
based on the gauge group $SU(2)$ with matter
in two vector multiplets and a singlet:
 $L^a : I = 1; H^a : I = 0; R : I = 0. $
   These `a' indices
transform with $i\e^{abc}$  and take the
values a=1,2,3.  Since the `a' indices are
real and since $\d_{ab}$ is an invariant tensor
of $SU(2)$, there is no difference when we raise and lower
these indices.

Without any good reason, we shall assume that the superpotential
does not contain a mass term for the $L$ field.

Since the `Higgs field' $H^a$ is in a real
representation of the gauge group,
it can have a mass term in the
superpotential.

Now we assume the following form for
the superpotential:
 \be W = g_1  L^a H^a R +
\fr{g_2 m}{2} H^a H^a + \fr{g_3}{4 m} [H^a
H^a ]^2
\ee
Note that renormalizability is not a property of this superpotential.

If $g_2 g_3 <0 $ , the Higgs field will
develop a VEV in its  `A' term that breaks
the gauge symmetry down to $U(1)$ while
leaving  supersymmetry unbroken.  The $L$
and $R$ fields develop no VEV.
Let us denote components as follows:
\be
L^a = A^a + \q^{\a}  \y^a_{\a} + \fr{1}{2} \q^2 F^a
\ee
\be
H^a = B^a + m u^a + \q^{\a}  \f^a_{\a} + \fr{1}{2} \q^2 G^a
\ee
\be
R  = A + \q^{\a}  \y_{\a} + \fr{1}{2} \q^2 F
\ee
We distinguish $ a=i,3$ where $1=1,2$.
Here we have included a shift by
the  VEV:
\be < B^a>_{\rm before\; shift} =
\d^{a 3} m \sqrt {\fr{- g_2} {g_3} }
\equiv
\d^{a 3} m h
\equiv
m u^a
\ee

Then the `F' term of the superpotential becomes:
 \[ W_F = \Bigl [ g_1  L^a (B^a + \d^{a 3} m h )  R +
\fr{g_2 m}{2}
(B^a + \d^{a 3} m h )(B^a +  \d^{a 3} m h )
\]
\be
+ \fr{g_3}{4 m}
[(B^a +  \d^{a 3} m h )
(B^a +  \d^{a 3} m h )]^2
\Bigr ]_F
\ee
In terms of components, this makes the following contribution
to the action:
\[
S_{{\rm Chiral}}
=
\int d^4 x W_F =
\int d^4 x \Bigl \{ g_1
( m h A^3 F +
 m h \y^{3 \a} \y_{\a}
+  m h F^3 A )
\]
\[
+ g_1 (A^a B^a F
+ \y^a \f^a A
+ F^a B^a A
+ A^a \f^a \y
+ A^a G^a A
+ \y^a B^a \y  )
\]
\be
+ {\rm terms\; involving\; H \; superfield \; only}
\Bigr \}
\ee
The essential point to note here is that there
is no mass term like
\be
m (  A^i F +
  \y^{i \a} \y_{\a}
+  F^i A )
\ee
for $i = 1,2$
which would give a mass to the $L^1$ and $L^2$ superfields.
They are massless after spontaneous breaking of the gauge symmetry.

The simplest composite operator (together
with a source $\F_{\a}$  that could develop
a supersymmetry  anomaly seems to be:
 \be S_{\rm	Composite}
	=	\int	d^{4}x	d^4	\q \Bigl	\{	\F^{\a}
{\ov D}^{\dot	\b} 	[ 		{\ov	L}^a
	{\ov	H}^{a}  ]
{\ov	D}_{\dot	\b}
	D_{\a}	 R	\Bigr	\}
\ee

After translation of the Higgs field, we
find the terms \[
 S_{\rm	Composite}
	=	m h \int	d^{4}x
\Bigl	\{
\c^{\a}
\Bigl	[
(\s^{\m
})^{\g	\dot	\b} \pa_{\m} {\ov A}^{ 3}
\s^{\n}_{\a	\dot	\b}
\pa_{\n} { \y}_{\g }
\]
\be
+
{\ov \y}^{3 \dot \b}
(\s^{\m })_{\a	\dot	\b} \pa_{\m} F
+ \cdots
\Bigr	] + \cdots
\Bigr	\}
\ee

The form of the supersymmetry anomaly that we would like to
calculate is
\be
\d \G
	=	m^4 \int	d^{4}x	 d^2 {\ov \q}
\F^{\a}	 c_{\a}
\sum_{i = 1,2}
{\ov L}^{i } {\ov L}^{i }=
	m^4 \int	d^{4}x
\c^{\a}	 c_{\a}
\sum_{i = 1,2}
{\ov A}^{i } {\ov A}^{i }
+ \cdots
\ee
Clearly at this point there are innumerable examples.  It may be
that even more structure is needed before non-zero examples of
supersymmetry anomalies can be found, or this may be non-zero itself.
I have not yet tried to calculate this example, but at least it
does not appear to be obviously zero, since it passes all the
tests mentioned above.

\section{Higher Spins}

Now we turn to another topic, which is
a fuller analysis of the cohomology of the BRS
operator defined by the supersymmetry invariance of
chiral multiplets of rigid $N=1,D=4$ supersymmetry.
The new result here \ci{dm}\ci{dmr} is that this cohomology space
contains potential anomalies in the renormalization of
fermionic superfields with all half-integer spins.
Formerly it had been shown that there were potential anomalies
for fermionic superfields with spin $\fr{1}{2}$ only \ci{prl}.

This may be very important for superstring theories, since such
higher spin multiplets necessarily occur in all such theories.

The result is that
there is an infinite set of states of the form
\[
{\cal {X} } _{\a_{1}...\a_{k_n}\dot \b_1...\dot \b_{k_n+g}}  =
\mbox{Sym}_{\dot \b_{1}...\dot \b_{k_n+g}} \int d^{4}x \, d^{2} \q
\Bl  \{\pa_{{\a}_1 \dot\b_1 }...\pa_{{\a}_{k_1} \dot{\b}_{k_1} }
 S_{a_1}...
\]
\be ...
\pa_{{\a}_{k_{(n-1)}+1} \dot\b_{k_{(n-1)}+1} }...
\pa_{{\a}_{k_n} \dot\b_{k_n} } S_{a_n} {\ov c}_{\dot \b_{k_n+1}}
{\ov c}_{\dot \b_{k_n+2}}...\ov c_{\dot \b_{k_n+g}}
\Br \}.
\ee
in the cohomology space of the Wess-Zumino model.
The corresponding complex conjugate expressions are obtained by converting
dotted into undotted indices and vice versa, and $S \ra \ov S, \; \ov c \ra c$.
By contraction and symmetrization of the undotted indices, we can decompose
$ {\cal{X} }$ into operators of the form
\be
{\cal {A}}_{(a,b)}={\cal {A}}_{(\a_1 \a_2 \cdots \a_{a}) ({\dot \b}_{1}
{\dot \b}_{2} \cdots
{\dot \b}_{b}) },
\la{irrep}
\ee
where $b=k_n+g$ and $k_n-a$ is even and greater than or equal to $0$, since
contractions always involve pairs of undotted indices. In particular, we are
interested in polynomials with ghost charge $g=1$, which correspond to
anomalies. For these objects, we find $b-a$ is  odd and positive. Therefore,
the operators ${\cal {A}}$  are spinors.

These objects could appear as anomalies in the renormalization of composite
operators with the same spin structure as the anomaly. To compute the anomalies
of a given such composite operator, a term
of the form
\be  S_{ \Y}=
\int \, d^{4}x \, d^{4}\q \, \left[ {
  \Y_{\a_{1}...\a_{a}\dot\b_1...\dot\b_{b}}}
{ \F^{\a_{1}...\a_{a}\dot \b_1...\dot \b_{b}}} \right]
\label{source}
\ee
would be introduced into the action.
Here $ \Y$ is a composite operator with ghost charge zero and
${\F^{\a_{1}...\a_{a}\dot \b_1...\dot \b_{b}}}$ is a chiral source superfield.
There is a matching between the indices of the anomaly and those
of the anomalous operator, because both must couple to the
source $\F$.
Now to compute the anomaly, some specific form for the
composite operator $ \Y$ would be chosen
and then the one particle irreducible
generating functional $\G$ including one vertex (\ref {source}) would
be calculated. If there is an anomaly, one would find that
the supersymmetric variation of this part of $\G$
would be of the form
\be
 \d \G= \k \int \, d^{4}x \, d^{2}\q \, \left[ {
  {\cal {A} }_{\a_{1}...\a_{a}\dot\b_1...\dot\b_{b}}}
\F^{\a_{1}...\a_{a}\dot \b_1...\dot \b_{b}} \right]
\la{ggg}
\ee
where $\k$ is a calculable coefficient.

Because all possible anomalies have half-integer spin (see discussion of
(\ref{irrep})), it follows that all operators which can be anomalous also
have half-integer spin. Generally, the entire class of spinor operators in
supersymmetric theories containing chiral matter can be anomalous.

We suspect that spontaneous breaking of the gauge symmetry will be
necessary for non-zero computation of superanomalies in
the higher spin cases just as in the spin one-half case.

Acknowledgments:  I would like to thank my collaborators Ruben
Minasian and Joachim Rahmfeld for many useful ideas, Ramzi
Khuri for some very useful discussions about integrals,
Heath Pois for many helpful remarks about
phenomenological models of supersymmetry, Chris Pope
for helping us to recognize the representations that arose in the
solution of the higher spin problems and Mike Duff for
his continued insistence that a supersymmetry anomaly be computed.

\end{document}